# Securing Refugee Identity: A Literature Review on Blockchain-based Smart Contract

**Dr. Md Taimur Ahad**
Associate Professor
Department of Computer Science and Engineering
Daffodil International University, Dhaka, Bangladesh
taimurahad.cse@diu.edu.bd

**Yousuf Rayhan Emon**
Teaching Assistant
Department of Computer Science and Engineering
Daffodil International University, Dhaka, Bangladesh
https://orcid.org/0009-0001-1088-0174
yousuf15-3220@diu.edu.bd

***Abstract:*** *- Identity documentation for refugees is a complex process and crucial for host nations. A secured identity management system ensures both security and the efficient provision of services for the host nation and the donor organizations. Realizing the benefits, a handful of studies enriched the blockchain-based security identification for refugees. The research studies presented the introductory, conceptual, and practical solution related to the blockchain-based smart contract. There is a common agreement in the studies that blockchain-based smart contract not only streamlines refugee identity verification but also safeguards against unauthorized entries. Since it is a technology as well, it has been essential to know the present status of the technology in the social context. In such a situation it becomes essential to review the existing research studies to provide insight for future studies. In this study, we reviewed current studies using a thematic approach. Our findings suggest researchers are more inclined to provide conceptual models as the models are important in advancing technology; however, the models need to be implemented for practical advances. However, the main contribution of this study is that this study gathers current efforts in smart contract-based refugee identity management. This study is important for the refugee host nations as well as for stakeholders. Knowledge gained from the study is expected to provide insight into how the technology can be developed using existing theory and implementation frameworks.*

## 1. Introduction:

Smart Contract (SC) has recently gained significant attention due to its decentralized, peer-to-peer transaction, distributed consensus, and anonymity properties. An SC is a set of programs that are self-verifying, self-executing, and tamper-resistant. SCs, with the integration of blockchain technology, are capable of securing information, such as identity preservation and identity management, in real-time at low cost and providing a greater degree of security ( Sohan et. al., 2022; Mohanta et al., 2018). The stored data is then controlled by smart contracts which define various access control policies, e.g., access parties, access times, etc. (Liu et al., 2019). The technical advantage of the SC is the technology makes the data in the identity management system safe and credible (Alsayed et al., 2019, Borse et al., 2019, Liu et al., 2017). SC effectively deals with privacy and security challenges on the one hand and enhances the decentralization and user control in transactions on Blockchain infrastructures on the other hand (Haddouti et al., 2019). In recent years, there have been attempts to introduce SC-based identity management solutions, which allow the user to take over control of his/her own identity (Liu et al., 2020). It sets itself up as the right person to tackle this challenge. That is mainly due to the use of SC cryptographic indicators, static recording, and background (Chauhan., 2021; Jindal et al., 2019).

United Nations defines a refugee as an individual who has been forced to flee from his or her country because of persecution, war, or violence. A refugee has a well-founded fear of persecution for reasons of race, religion, nationality, political opinion, or membership in a particular social group. Most likely, the study cannot return home or are afraid to do so. War and ethnic, tribal, and religious violence are leading causes of refugees fleeing their countries (unrefugees.org). However, identity management is important for the refugees. The identity of refugees is important for navigating the various identities available to them, consciously weighing the benefits and constraints of different statuses to maximize access to services, eligibility for employment, and mobility (Shoemaker et al., 2019). Moreover, donor organizations, such as the United Nations (UN) is seeking ways to give refugees control over their own individual identities and greater forms of self-determination within contexts of assistance and protection (Franke et al., 2022).

In general, refugee data are collected by the host nations, UN, or donor agencies, Usually, refugees are typically not able to exercise the data that are collected about them (Shoemaker et al., 2019). In refugee identity management, security and privacy are two main features (Alsayed et al., 2019). Moreover, the trustworthiness of humans and institutes who are responsible for controlling the entire activity is also impacting the security mechanism (Alsayed et al., 2019). However, the host nation has to spend a lot of time and assets to document the refugee's identification. Moreover, the existing refugee identity system process is error-prone, inefficient, and resource-intensive (Habib et al., 2023). In such a situation, there is a call for a new identity management system to provide a more promising secure platform (Lim et al., 2018).

In such a situation, SC as a distributed ledger technology positions itself as a suitable candidate to address the challenges associated with the refugee system (Omar et al., 2018). With the growing concerns regarding personal information, as well as with the General Data Protection Regulation (GDPR), SC mainly includes identity preservation, authentication, and management. For example, (Liu et al., 2017) pointed out that smart contracts can write rules to ensure the reliability of identity information, such as who can access the identity. Moreover, using SC, users can maintain and manage their identities associated with certain attributes, accomplishing the self-managing ability.

Despite the fact numerous studies suggested implementing smart contract-based refugee identification, the initial literature survey suggests that most of the studies on blockchain-based refugee identification have yet not been implemented (Kolhatkar et al., 2023, Sohan, et al., 2022). Most studies focused on providing a conceptual model (Ngo et al., 2023). Most of the studies are conducted in US, India, and China (Ngo et al., 2023). However, Bangladesh, which hosts around one million refugees was not attracted by researchers for blockchain-based refugee identity management. Though some studies provide a snapshot in this regard (Habib et al., 2023), the study does a real process of refugee identity management. This creates a gap in understanding how different stakeholders in refugee management will interact in an SC-based distributed ledger. Following the shortcoming, this study a software development methodology and provides a use case of SC for refugee identity management. The study is important, as it provides essential knowledge on how the smart contract-based security model ensures the authenticity of identity documents and conserves fundamental rights for refugees. In constructing the model, a secure environment for identification is demonstrated, where the authorized controller, the third party's data request, reviews and assists refugees' functions.

# 2. Literature review:

In the literature review, we first discuss our process of finding literature, then we present the literature.

## A. Exploring Blockchain Technology: Process of Literature Review

Diving into blockchain technology requires a systematic approach like Systematic Reviews and Meta-Analyses. This literature review aims to uncover the landscape of blockchain research, focusing on key aspects such as applications, consensus mechanisms, scalability solutions, smart contracts, and integration with traditional industries. The goal is to identify empirical research articles, presenting concrete experimental findings over theoretical conjecture.

**Article Identification**

Identifying the Right Articles With around 34 carefully selected papers forming the bedrock of this exploration, the methodology echoes the thoroughness of the PRISMA guidelines. Blockchain technology is complex, but these papers provide insight into its intricacies. The process of identifying relevant articles begins by searching through influential digital libraries, such as IEEE Xplore, ACM Digital Library, and Google Scholar, as shown in Figure 1. These platforms serve as starting points for uncovering articles that showcase the potential of blockchain.

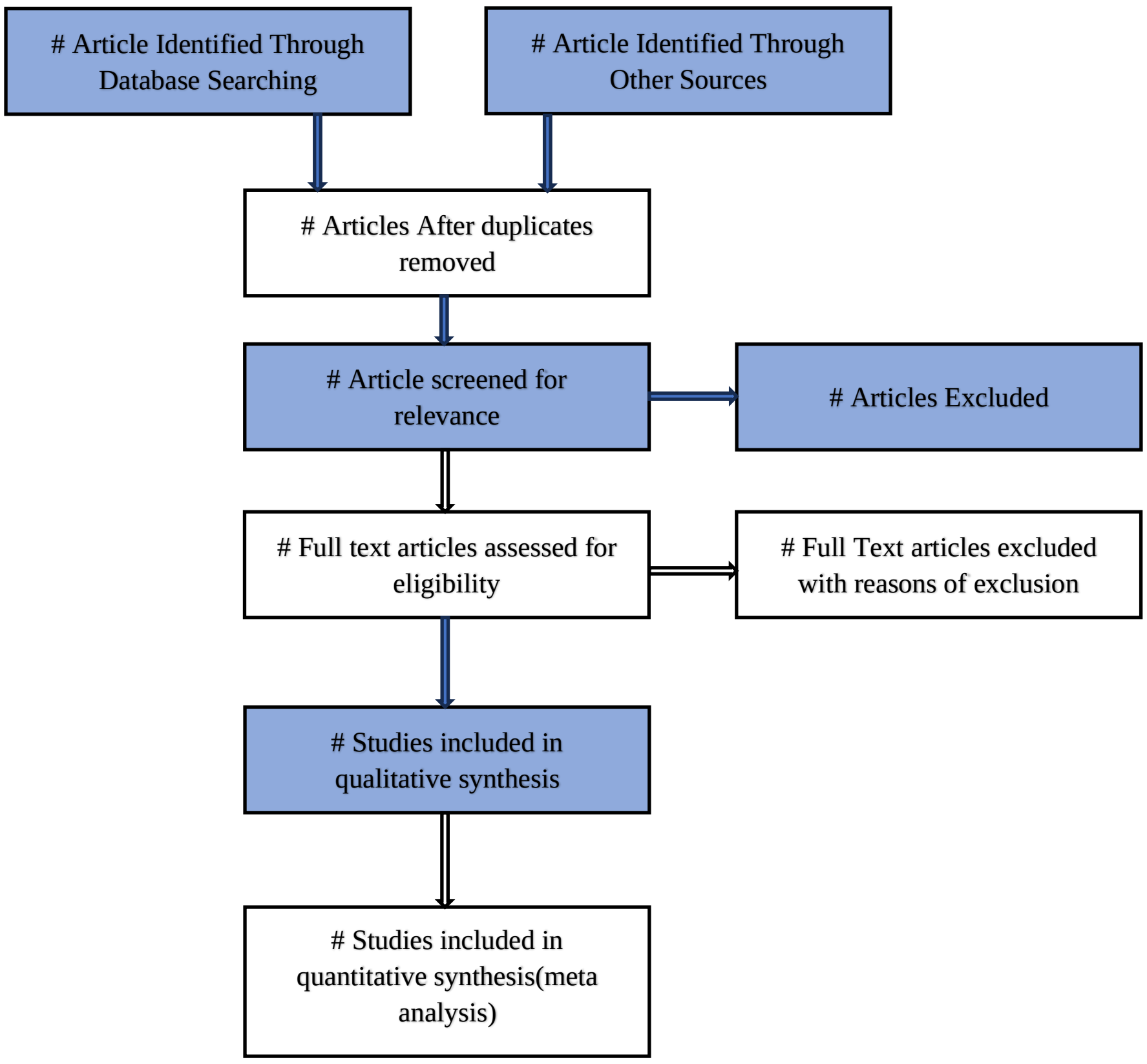

Figure 1: Article Identification Process

**Article Selection**

The process of article selection for the blockchain-based literature review involved an accurate three-stage screening approach characterized by a set of predefined inclusion and exclusion criteria (see Figure 2). After eliminating duplicates from multiple databases, articles underwent screenings based on titles, abstracts, and full articles. The principal investigator meticulously carried out this entire screening process.

To meet the specified inclusion criteria, articles were required to fulfil the following conditions:

- The study had to be original research articles, formally published in peer-reviewed journals, and accessible in full-text through the University's resources.
- It needed to involve the utilization of various forms of blockchain technology, including its applications, scalability solutions, and smart contracts.
- Publications were restricted to those written in the English language.
- Implementing blockchain technology across industries, exploring its potential integration with traditional sectors.
- Specifically, articles were limited to those published between 2017 and 2023 to emphasize recent advancements and contemporary methodologies in the blockchain domain.

Exclusion criteria encompassed the following:

- Short communications or case reports
- Articles with unclear descriptions of blockchain implementations or methodologies
- Studies that lacked proper validation processes

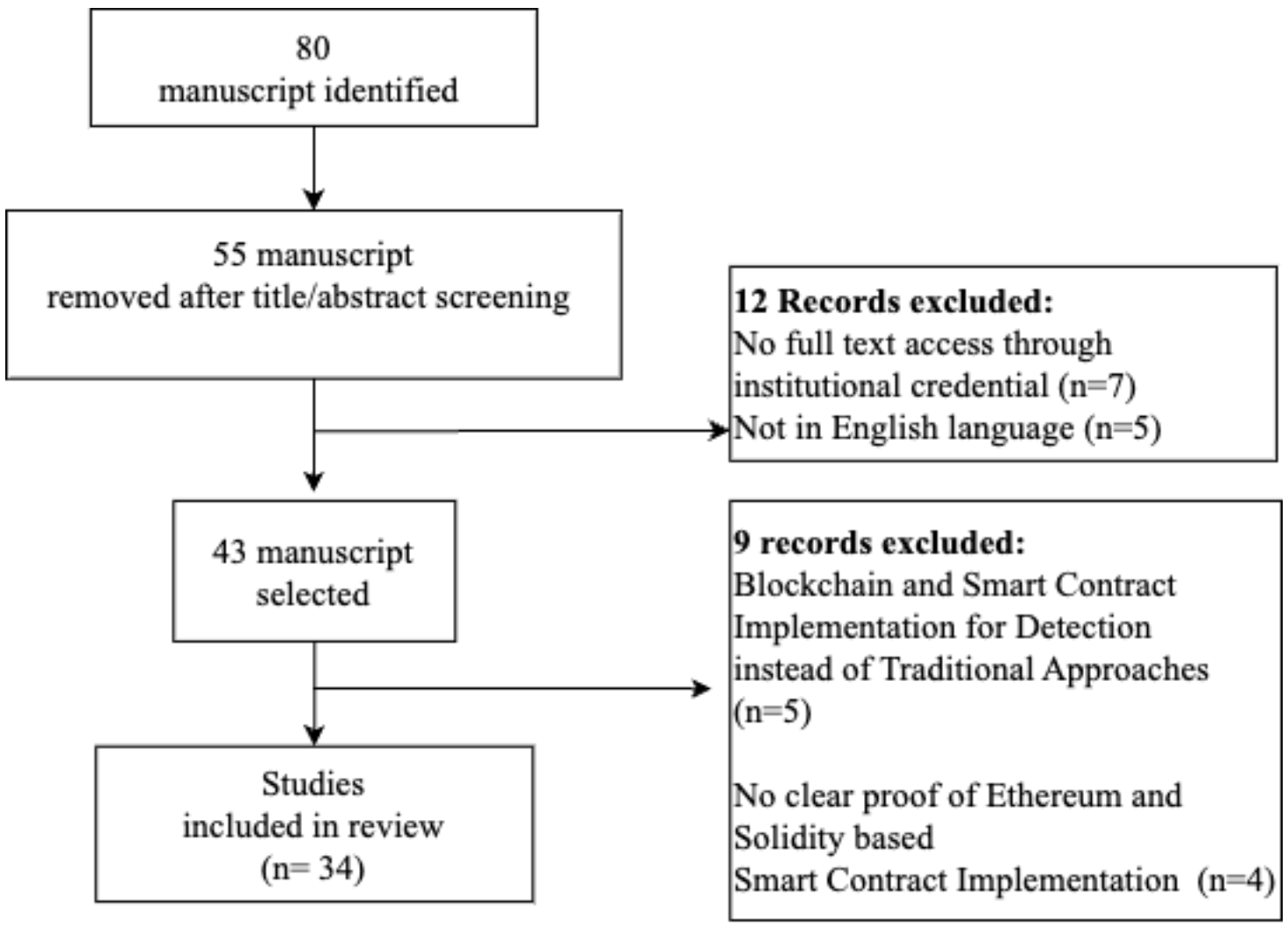

Figure 2: Article selection process

## B. Literature Review

The literature review suggests conceptual, solution proposal, validation research, evaluation research, opinion papers, and experiential papers.

Ngo et al. (2023) conducted a comprehensive review of 361 blockchain papers. These papers came from major databases and were written over the last 13 years. Their study was set apart by focusing on how often other studies mentioned or "cited" these papers. This gives a good idea about which papers had the most impact. Taherdoost (2023) has narrowed his focus on examining the role of blockchain technology in facilitating smart contracts. He reviewed 252 papers from the past ten years, trying to understand where smart contracts stand today. Besides just giving an overview, he also pointed out where more research is needed and the challenges people might face when working on these topics.

Going deeper into real-world uses, Kordestani et al., (2023) focused on the medical world. They wanted to see how smart contracts in blockchain could help this industry, especially when it comes to the big problem of fake medicines. Their findings showed the unique ways smart contracts can be used and how they can stop these fake medicines from reaching patients. They also gave ideas on what researchers and business people should consider next. Agrawal et al., (2023) thought about how to get different groups to work together using blockchain. They created a new way to share resources with the help of smart contracts. This new plan ensures that the resources shared are genuine and helps when there's either too much or too little of something in supply chains.

Blockchain is a technology that can help in many areas, especially for people moving from one country to another because of troubles in their own country. Connolly et al., (2023) looked at how blockchain can help these people, mainly by giving them a kind of online ID. They studied big projects like the one by the World Food Programme in Jordan and another for the Rohingya people. They found that while blockchain can be suitable for these people, it can also harm them by removing their privacy if not used correctly. Habib et al., (2023) showed a better way to use blockchain to manage IDs. They ensured that with this system, people who come to a new country for safety couldn't just become citizens. They also talked about how this system keeps data safe.

Visvizi et al., (2023) discussed places where people go in hopes of getting safety in a new country. These places sometimes could be better to live in. So, they suggested a tool called "responsible wallet" to help make things better there. Kolhatkar et al., (2023) used blockchain for something different: to help when natural disasters happen. They made a system where different groups, like aid organizations and governments, can work together better. This system has lists of missing people, maps of camps, and info on aid distribution.

Arafat et al., (2022) find out the major problem like fake IDs. They made a blockchain system to check if IDs are real or not. They also showed how this system can help refugees get what they need from countries helping them. Franke et al., (2022) shared findings on a project by the UNHCR. They said that while blockchain can help refugees manage their IDs, it's also reducing them to just numbers in a system. It's important to remember that there's more to a person's identity than just data.

Demir et al., (2020) discussed how blockchain can change our thinking about trust, especially when bad things happen, like natural disasters. They made a new system that helps all the groups involved, from donors to victims, work together better. Shoemaker et al., (2019) studied how personal data of refugees in some countries is managed. They found many challenges, but refugees also played a big part in accessing their needs. These studies show that blockchain can help in many areas, from managing IDs to helping after natural disasters. But we need to use it correctly to help people truly.

Traditional identity systems store information in one place, which can be risky. If something goes wrong with that central system, user data might be exposed. Yang et al., (2020) noticed these risks and discussed how older systems have issues like data leaks and other threats. They found that using a blockchain can solve many of these problems. Liu et al., (2017) also worked on a decentralized identity system that uses the blockchain. This kind of system is safer because it doesn't have one weak point. Plus, blockchain technology makes sure the data is trustworthy. Having control over your own identity online is a big deal. Song et al., (2022) talked about using blockchain to make a system where people can manage their digital identity independently. This type of system gives more power to the user and is called a self-sovereign identity. Liu et al., (2020) also discussed how people should be able to control their digital identity and reviewed a lot of research on this topic. Lim et al., (2018) looked into different ways to manage identities using the blockchain and tried to make it even safer.

The traditional systems sometimes use personal details, like our fingerprints or facial features, to check who we are. But these systems can be risky. Liu et al., (2019) talked about the dangers of these older systems. They suggested a new way to use personal details safely with the blockchain. This way, the user can decide who can see their details and for what purpose. Zhao et al., (2019) also mentioned the risks of the old system and suggested a new system that gives users more control. Arafat et al., (2022) worked on a unique system for helping refugees. This system uses the blockchain to make sure refugee documents are actual and not fake. It also helps with other tasks like providing aid to refugees. Madon et al., (2021) discussed how the UNHCR, a big organization that helps refugees, can use digital tools to do its job better. These tools can help solve problems and ensure refugees get the help they need. Lee et al. (2017) proposed a blockchain-based system for recovering lost identity information. They made a system where a person's friends can help them recover their details if they lose them. Omar et al., (2019) introduced a new system that connects the Internet of Things (IoT) with the blockchain. This system helps create and move digital identities between networks. Chauhan et al., (2021) worked on a different idea. They used the blockchain to manage patents, like certificates for new ideas. Their system helps people register and transfer ownership of these ideas.

Table 1: Research matrix of blockchain-based study

| Study | Domain | Study Type | Smart Contract build? | Advantages of this study | Limitations of this Study |
|---|---|---|---|---|---|
| Ngo et al., (2023) | Blockchain in academic research | Literature review | Not specified | Comprehensive view of articles, analysis citations, most significant number of articles studied. | Not explicitly mentioned. |
| Taherdoost, H. (2023). | Smart contracts in blockchain | Literature review | Not specified | Overview of smart contracts' status and significance. | Identified gaps and challenges in the literature. |
| Kordestani et al., (2023) | Pharmaceutical blockchains | Systematic literature review | Not specified | Identifies characteristics and roles of smart contracts in the pharmaceutical supply chain. | Not explicitly mentioned. |
| Agrawal et al., (2023). | Resource sharing using blockchain | Implementation framework | Yes, for resource sharing. | Ensures quality and data authenticity, useful for effective resource utilization. | Not explicitly mentioned. |
| Connolly et al., (2023) | Protecting human rights of migrants and refugees using blockchain | Survey and case study | Not mentioned | Shows how blockchain can empower vulnerable individuals. | Potential human rights risks identified. |
| Habib et al., (2023) | Identity management for refugees | Demonstration | Yes, for identity management. | Enhanced effectiveness of identity management. | Not explicitly mentioned. |
| Visvizi et al., (2023) | Support for irregular migrants and refugees | Proposal | Proposed "responsible wallet" tool. | Addresses challenges faced by irregular migrants. | Not explicitly mentioned. |
| Kolhatkar et al., | Disaster management | Implementation | Yes, for various disaster | Streamlines disaster | Not explicitly mentioned. |
| (2023) | | | management aspects. | response, enhances collaboration and efficiency. | |
| Arafat et al., (2022) | Identity documents' authenticity for refugees | Proposal | Yes, for identity verification and aid management. | Ensures authenticity of IDs, conserves fundamental rights. | Not explicitly mentioned. |
| Franke et al., (2022) | Identity management for refugees | Survey | Not mentioned | Refugees can control their identity. | Misunderstanding of identity, displacement of relational qualities. |
| Demir et al., (2020) | Disaster relief and aid | Proposal | Yes, for disaster relief. | Enhanced processes, promotes willingness to help. | Not explicitly mentioned. |
| Shoemaker et al., (2019) | Identity management for refugees | Survey | Not mentioned | Insight into challenges of existing humanitarian identity systems. | Not explicitly mentioned. |
| Yang et al., (2020) | Digital Identity Management System (DIMS) in blockchain | Implementation | Yes, for claim identity model improvement | Enhanced attribute privacy and wider application scope compared with the prior model | Challenges of transparency and privacy in blockchain |
| Liu et al., (2017) | Decentralized Identity Management on Blockchain | Implementation | Yes, for system rules and user information reliability | Secure identity and reputation management on the Internet; data safety and credibility | Not explicitly mentioned. |
| Song et al., (2022) | Digital Identity Verification and | Implementation | Yes, for ZKP verifications and digital | Efficient, safe system; | Not explicitly mentioned. |
| | Management Systems (DIVMS) of blockchain-based verifiable certificates (VC) | | signature of IDP | improved privacy of blockchain-based VC; overcome single point failure; superior to some existing blockchain-based DIVMSs | |
| Liu et al., (2020) | Blockchain-based Identity Management | Literature Review | No (Review of existing systems) | In-depth analysis of blockchain-based identity management; identification of research gaps and opportunities based on analysis | Limited to papers and patents published between May 2017 and January 2020. |
| Lim et al., (2018) | Internet Identity Protocol | Survey/Literature Review | Not mentioned | New solution for identity protocol, Disrupted existing identity management and authentication solutions. | Not explicitly mentioned. |
| Liu et al., (2019) | Biometrics for Identity Authentication | Implementation | Yes, for controlling access to biometrical data on Ethereum | Secure and privacy-preserving approach for biometrics-based identity management, flexibility in access control policies. | Not explicitly mentioned. |
| Zhao et al., (2019) | Decentralized Identity Management | Implementation | Yes, for decentralized identity management | Decentralized system, user control over identity information, credible identity in decentralized environment, attribute | Not explicitly mentioned. |
| | | | | reputation. | |
| Madon et al., (2021) | Digital Platforms in Humanitarian Contexts | Empirical Study | Not mentioned | Insights into UNHCR's data transformation strategy, understanding of platform governance-related tensions, issues of exclusion and vulnerability in refugee contexts. | Not explicitly mentioned. |
| Lee et al., (2017) | Secure Firmware Update for IoT Devices | Implementation | Not mentioned | Secure and reliable firmware update for embedded devices in IoT environments, protection against firmware tampering and vulnerabilities. | Not explicitly mentioned. |
| Omar et al., (2018) | IoT Devices Identity Management | Implementation | Yes, for identity creation and transfer of ownership | Unique and global identity for IoT devices, identity portability, and ownership transfer features in semi-decentralized framework. | Not explicitly mentioned. |
| Chauhan et al., (2021) | IoT Devices Identity Management (Note: This abstract seems to be a rephrasing of the previous one) | Implementation | Not mentioned | Focus on patent management, device identity maintenance throughout its lifecycle, registrar and management smart contracts. | Not explicitly mentioned. |
| Sohan et al., (2022) | Blockchain with IoT in pharmaceutical industries | Conceptual model | Not specified | Provides a secure pharmaceutical product delivery. Ensures traceability of | Not mentioned. |
| | | | | drugs in a simple, less complex manner. | |
| Jindal et al., (2019) | Secure Energy Trading in V2G Environment | Proposed framework | Not specified in abstract | Securing energy trading via blockchain, reducing latency and overhead, enhancing smart transport network throughput, effective for EV-CS transactions. | Not mentioned. |
| Dunphy et al. (2018) | DLT-based Identity Management (IdM) | Evaluation | Not specified | Introducing DLT-based IdM, evaluating uPort, ShoCard, Sovrin, and analyzing DLT's role in decentralized identity. | Not mentioned. |
| Haddouti et al., (2019) | Identity Management | Comparative Analysis | Not specified | Analyzes uPort, Sovrin, and ShoCard, evaluating their digital identity features for concise, reader-friendly presentation. | Not mentioned. |
| Borse et al., (2019) | Identity Management | Research Proposal | Not specified | Introduces self-sovereign identity in blockchain using zero-knowledge protocol for selective user attribute anonymity. | Not mentioned |
| Alsayed Kassem et al., (2019) | Identity Management | Research Proposal | Yes (DNS-IdM) | Proposes DNS-IdM, a smart contract system for self-sovereign identity | Flaws still exist in the decentralized system |
| | | | | management with enhanced security and privacy | |
| Sultana et al., (2020) | IoT Communication | Research Proposal | Yes (ACC, RC, JC) | Introduces a blockchain system for IoT data sharing and access, utilizing smart contracts for control, misbehavior detection, and efficient gas usage. | Not mentioned |
| Ullah & Al-Turjman., (2023) | Smart Real Estate | Systematic Review | Not Specified | Comprehensive review of blockchain smart contracts in real estate, proposing a six-layered adoption framework and aligning with Industry 4.0 standards | Not mentioned |
| Hunt et al., (2021) | Humanitarian Operations Management (HOM) | Systematic Review | Not Specified | Reviewing blockchain in healthcare operations: focus areas, benefits, barriers, and practical advancements. | Most works on blockchain in HOM are untested in the field. Limited empirical evidence on blockchain's capabilities in HOM. |
| Mohanta et al., (2018) | Blockchain and Smart Contracts | Comprehensive Review | Not Specified | Examining smart contract mechanics, applications, benefits, cost-efficiency, and security in blockchain technology. | Challenges in implementing smart contracts in real-world scenarios are not properly discussed. |

## 3. Inference from current research studies

Discussion of Findings from Literature from Table 1, it is understandable that, though identity management has been highlighted as very important, however, most of the studies are at the introductory stage (Ngo et al., 2023). The introductory studies mainly discussed the advantages and implications of blockchain in identity management. Another main contribution to the body of knowledge is through a conceptual framework. Conceptual models and frameworks are also important in advancing technology; however, the methods are yet to be implemented. Another theme is providing a literature review. Literature reviews are essential in formulating and informing current trends. As a technology, blockchain-based identity management has to be implemented for business advancement, which we found rare. With implementation, the advantages of technology can be enjoyed.

Moreover, business-to-business, industry-to-industry, and country-to-country identity management differs. Hence, more use cases on blockchain-based identity management are required. Implementation and use cases will mature the technology as IT practitioners and software houses will contribute with their technical strength.

## 4. Limitations

Like many other research studies, this research also inherits some limitations. Firstly, since we have a selection criterion for a research paper for inclusion in this study, some impactful research in this domain should have been included in this research. However, to minimize the limitation, we randomly attempted to search for studies that are accepted in reputed Q1 journals. Secondly, because of resource constraints, we could not obtain paid articles. We only included papers available in scholar, academia, and other free respiratory. In this situation, we tried to get the papers from our peers. Since we observed the number of conceptual studies is more than the implemented paper, the distribution of the literature review is not balanced. A balanced literature might enrich the review.

## 5. Direction for Future Research

The study of blockchain-based identity management has provided some basis for future research. These are as follows:

- The blockchain-based identity management framework needs to be tested in different industries.
- Future research should include advanced identity management such as biometrics, figure prints, and voice recognition-based identity management in blockchain technology.
- Security perspective factors should be revealed to extend our knowledge of how blockchain can secure various industry perspectives. This will enable us to understand more about the capabilities of blockchain.
- Future research should also include how more easy technology can be included in the blockchain.

# 6. Conclusion

This study represents a review of blockchain-based identity management screening where we describe the article selection process and current efforts in blockchain-based security inference from the literature review. Furthermore, we considered the actual process of refugee identity management rather than providing an unrealistic process of refugee identity management. This is one of the few studies incorporating the latest authentication, such as biometrics and Irish scans. We have focused on the strengths and weaknesses of some studies based on their approach and models. Lastly, we have provided the future research direction on blockchain-based identity management.